# Parity-violation in bouncing cosmology

**Mian Zhu**[a,b] **and Yong Cai**[a]

[a]*School of Physics and Microelectronics, Zhengzhou University,
Zhengzhou, Henan 450001, China*

[b]*Faculty of Physics, Astronomy and Applied Computer Science, Jagiellonian University,
30-348 Krakow, Poland*

*E-mail:* mzhuan@connect.ust.hk, yongcai_phy@outlook.com

ABSTRACT: We investigate the possibility of the enhancement of parity-violation signal in bouncing cosmology. Specifically, we are interested in deciding which phase should generate the most significant parity-violation signals. We find that the dominant contribution comes from the bouncing phase, while the contraction phase has a smaller contribution. Therefore, bouncing cosmology can enhance the parity-violation signals during the bouncing phase. Moreover, since the bouncing phase has the highest energy scale in bouncing cosmology, we can also probe new physics at this scale by studying the parity-violation effect.

## Contents

## 1 Introduction

The primordial gravitational waves (GWs) might encode rich information about the very early universe, which may help distinguish between different scenarios of the primordial universe, including inflation and its alternatives. Chirality is a distinct characteristic of GWs, which could be manifested in parity-violating theories of gravity. Recently, it is found that the polarization data of Planck and WMAP [1–3] may be a hint of parity-violating physics in the cosmic microwave background, though further confirmations are required. The explorations of parity-violating primordial GWs have aroused a lot of interest, see e.g. [4–39], see also [40–47].

In single-field slow-roll inflation models where inflaton is non-minimally coupled to a parity-violating term, such as the gravitational Chern-Simons (gCS) term [48, 49], the effect of parity-violation should be suppressed by the slow-roll condition. However, in the modifications or alternatives to the single field slow-roll inflation, the slow-roll condition could be violated at least for some moment. As a result, the effect of parity-violation could be enhanced due to the dynamical coupling of the scalar field to the parity-violating term, see e.g. [50] for the enhanced parity-violating GWs caused by violation of the null energy condition (NEC) [51] during inflation. Therefore, observations of a parity-violating GW background might provide us with a new way to identify physics beyond single-field slow-roll inflation.

Bouncing cosmology as a possible solution to the initial cosmological singularity problem of inflation and the Big Bang cosmology has attracted a lot of interest [52–79]. In the bouncing scenario, the universe originates from a contracting phase and enters an expanding phase after going through a non-singular bounce, where the NEC is violated. One important issue is the ghost and gradient instabilities in the bouncing phase, which is a generic feature in a large class of theories [80–82]. To acquire a healthy bouncing model, new physics effective at bouncing phase is introduced [83–100].

In principle we may explore the new physics by studying their phenomenological predictions. Unfortunately, the signals from the new physics, which is generically effective only in the bouncing phase, are suppressed by the short duration of the bounce. For example, in [95], it is found that the new physics at the bouncing phase has negligible contribution to the power spectrum. Consequently, in many studies of non-singular cosmology [101–108], the signals from the bouncing phase are small, the main phenomenological contribution comes from the contraction phase, so it's difficult to probe the new physics in bouncing phase.[1] Specifically, in previous literature addressing the parity-violation effect in bouncing cosmology [10], the bouncing phase is directly replaced by a simple junction condition so there is only a contraction and an expansion in their scenario.

It is then interesting to study if the parity-violation effect could be generated in bouncing cosmology, especially during the bouncing phase. Intuitively, the derivative of the scalar field is non-trivial around the bouncing phase, which may be able to amplify the effect of parity-violation, as long as the scalar field is non-minimally coupled to a parity-violating term. Additionally, the effective sound horizon of the primordial GW mode could also be nontrivial during the bouncing phase,[2] especially when chirality of the GW mode is considered. Therefore, we expect the non-trivial parity-violation signals to come from the bouncing phase.

In this paper, we investigate the parity-violation effect in a toy bouncing model, where the source term is taken to be a gCS action coupled to the scalar field. We are especially interested in the following question: which phase, the contraction or the bouncing phase, contributes to the enhancement of the parity-violation effect dominantly? As we will see in section 3.3, the bouncing phase can generate non-trivial parity-violation signals, while the contraction phase has negligible effect. Moreover, the enhancement is sensitive to the detailed physics during the bouncing phase, so in principle, we can probe the new physics during bouncing through parity-violation. Therefore, our result is twofold: we can not only explain the possibly observed parity-violation signal in the framework of bouncing cosmology, but also provide a possible way to probe new physics at the bouncing phase by studying their imprint on parity-violation signals.

The paper is organized as follows. In section 2 we briefly introduce our model. After the basic formalism for tensor perturbation in 3.1, we numerically evaluate the dynamics of tensor perturbation in section 3.2 and the parity-violation signal in section 3.3. We comment on some conceptual issues about our result in section 3.4 and explain our numerical result in a semi-analytical way in section 3.5. From the semi-analytical argument, we find that our numerical result should be qualitatively valid for a large variety of bouncing

---

[1]Some counterexamples comes from the quantum bounce models [109–111]. However, this is beyond our scope since we consider purely classical bouncing cosmology.

[2]The bouncing phase is defined by $dH/dt \geq 0$, where $H$ is the Hubble parameter.

– 2 –

models, although the numerics are taken in a toy bouncing model. We finally conclude in section 4.

Throughout this paper, we take the sign of the metric to be $(-,+,+,+)$. We will take $\hbar = 1$, $c = 1$, $M_p^2 = (8\pi G)^{-1} = 1$, so that all quantities are in Planck units. The canonical kinetic term is defined as $X \equiv -\nabla_\mu \phi \nabla^\mu \phi /2$, such that $X = \dot\phi^2/2$ at the background level.

## 2 Model

### 2.1 Action

We take the action to be

$$S = \int d^4x \sqrt{-g} \left[ \frac{M_p^2}{2} R + \mathcal{L}_H + \mathcal{L}_G + \mathcal{L}_{HE} \right]. \tag{2.1}$$

The term $\mathcal{L}_H$ is responsible for setting the background evolution, where we set

$$\mathcal{L}_H = M_p^2 f_1(\phi) X + f_2(\phi) X^2 - M_p^4 V(\phi), \tag{2.2}$$

which is eligible for the background dynamics. In the next section 2.2, we will use specific coupling functions $f_1$ and $f_2$ to construct a cosmological bouncing model.

The $\mathcal{L}_G$ term is the gravitational CS term, with

$$\mathcal{L}_G = \frac{f_3(\phi)}{8} R \wedge R = \frac{f_3(\phi)}{8} \epsilon^{\alpha\beta\rho\sigma} R_{\alpha\beta\mu\nu} R_{\rho\sigma}{}^{\mu\nu}, \tag{2.3}$$

and $\epsilon^{\alpha\beta\rho\sigma}$ to be four-dimensional Levi-Civita symbol with $\epsilon^{0123} = -1/\sqrt{-g}$.

Finally, the term $\mathcal{L}_{HE}$ represents the action effective at some high energy scale. Since there will be ghost or gradient instability problems in the generic bouncing models [80, 81, 104], such terms are obligated to eliminate such instabilities. We will discuss in details of this term in section 2.3.

We mention that in (2.1) we scale the scalar field $\phi$ to be dimensionless so that the coupling functions $f_i$ are dimensionless.

### 2.2 Bouncing background

It is well-known that the gCS term will not contribute to the background dynamics. We shall assume that the correction term $\mathcal{L}_{HE}$ also satisfies this criterion. Therefore, Friedmann's equations are totally determined by the Einstein-Hilbert action and the $\mathcal{L}_H$ term. In a flat FLRW background

$$ds^2 = -dt^2 + a^2(t) d\vec{x}^2, \tag{2.4}$$

we have

$$3M_p^2 H^2 = \frac{M_p^2}{2} f_1 \dot\phi^2 + \frac{3}{4} f_2 \dot\phi^4 + M_p^4 V(\phi), \tag{2.5}$$

$$-2M_p^2 \dot H = M_p^2 f_1 \dot\phi^2 + f_2 \dot\phi^4, \tag{2.6}$$

or in terms of the scalar field $\phi$:

$$\left(M_p^2 f_1 + 3\beta\dot{\phi}^2\right)\ddot{\phi} + 3H\dot{\phi}\left(M_p^2 f_1 + \beta\dot{\phi}^2\right) + M_p^4 \frac{dV}{d\phi} + \frac{M_p^2}{2}\frac{df_1}{d\phi}\dot{\phi}^2 = 0 \; . \tag{2.7}$$

Now we choose a similar ansatz as that from [90, 95]:

$$f_1(\phi) = 1 - \frac{g}{\cosh\omega_1\phi} \, , \; f_2 = \beta \equiv \text{const} \, , \; V(\phi) = -\frac{V_0}{\cosh\omega_V\phi} \, , \tag{2.8}$$

where the background dynamics are well-studied. In the initial state of the universe where $\dot{\phi} \to 0$ and $\phi \to -\infty$, the universe undergoes an Ekpyrotic contraction [112] (also known as the slow contraction, see e.g. [113])

$$\phi \simeq -\frac{1}{\omega_V}\ln\frac{\omega_V^4 V_0 t^2}{\omega_V^2 - 6} \, , \; a(t) = a_-\left(\frac{t-t_c}{t_- - t_c}\right)^{\frac{2}{\omega_V^2}} \; . \tag{2.9}$$

The Ekpyrotic phase makes us free from conceptual issues of bouncing cosmology [114], at the cost of requiring $\omega_V^2 > 6$. Note that we set $t = 0$ to be the bouncing point, i.e. the stage where the scale factor is minimal, so need an integration constant $t_c$ to correctly describe $a$. We also use the minus sign to denote the end of the Ekpyrotic phase, e.g. $a_-$ is the scale factor at the end of the Ekpyrotic contraction.

When $|\phi| \to 1$, the hyperbolic function approaches 1, and if we take $g > 1$, the $f_1 X$ term inverses sign and NEC can be violated. The non-singular bounce phase starts when the NEC is violated, and the universe transit from contraction to expansion. The dynamics during the bouncing phase are generically complicated, but for a short bounce, i.e. bouncing phase with short enough time, the following parameterization can be valid

$$H = \gamma M_p^2 t \, , \; \gamma = \text{const.} > 0 \; \to \; a = a_0 e^{\frac{1}{2}\gamma M_p^2 t^2} \, , \tag{2.10}$$

where we have set $a_0 = a(0)$, which is the scale factor at the bouncing point.

After the bouncing phase, the universe comes to an expansion phase, where the scale factor behaves as

$$a(t) = a_+\left(\frac{t-t_e}{t_+ - t_e}\right)^{\frac{1}{3}} , \quad H(t) = \frac{1}{3(t-t_e)} \, , \tag{2.11}$$

where we similarly use the "+" sign to denote the end of the bouncing phase, and $t_e$ is another integration constant.

We shall comment more on the expansion phase. Notice that, the factor $aH$ from (2.11) is proportional to $(t-t_e)^{-\frac{2}{3}}$. Hence, for any wave mode that is initially sub-horizon at $t = t_+$, it will remain sub-horizon in the whole expansion phase. This is in contrast with our general belief that, the primordial perturbation should leave the horizon in the expansion phase (like inflation) and freeze in, and re-enter the horizon in a later stage to set the initial condition for structure formation.

However, the parity-violation signal is highly dependent on the subsequent expansion phase after the bounce. In this paper, we want to compare the induction of parity-violation between the contraction phase and the bouncing phase, so we wish to get a result independent of the subsequent expansion phase. Unfortunately, we do not have a precise way to define when the bouncing phase ends, so it is hard to directly get the parity-violation status at the end of the bouncing phase. The advantage of our expansion phase (2.11) is that the wave mode of interests will always be in the sub-horizon region. Thus, their dynamics can be approximately described by the harmonic oscillator equation $u_k'' + c_T^2 k^2 u_k = 0$, whose general solution is simply

$$u_k \simeq u_{k,+} e^{ik\tau} + u_{k,-} e^{-ik\tau} \ . \tag{2.12}$$

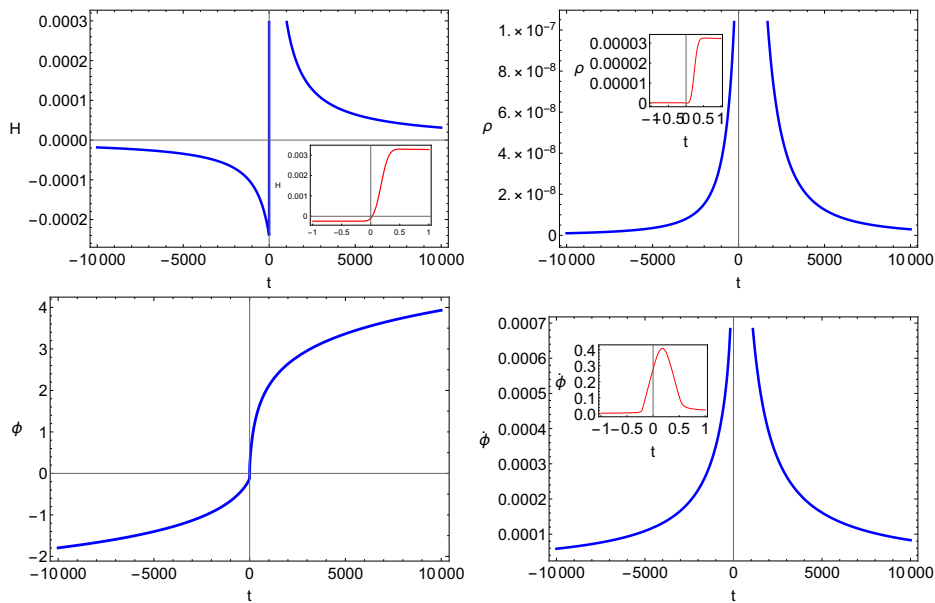

**Figure 1**. The background dynamics with the specific parameters (2.13). The upper channel shows the evolution of the Hubble parameter and background energy density, while the lower channel shows the dynamics of the scalar field $\phi$. The bouncing phase happens at around $t=0$ where $H$ quickly transfers from negative to positive.

The information of parity-violation status when bounce ends is encoded in the function $u_{k,\pm}$, and we see that the expansion phase only changes their relative phase. Thus, we may alternatively get the physics of parity-violation at the end of the bouncing phase, by tracing the statistical property of tensor perturbation during the expansion phase. We shall elaborate more about this point in section 3.2.

We depict the background dynamics in figure 1, where we've adopted the following parameters

$$g = 1.5 \ , \ \beta = 2 \ , \ V_0 = 10^{-7} \ , \ \omega_1 = 10 \ , \ \omega_V = \sqrt{10} \ . \tag{2.13}$$

### 2.3 The effective action on high energy scale

As shown in figure 1, the background energy density at the bouncing phase is much higher than the other phases. Hence, it is natural to introduce some actions effective only at a high energy scale to eliminate the instability problem. In the context of effective field theory (EFT) of non-singular cosmology [83, 84], certain EFT operators such as $R^{(3)}\delta g^{00}$ can help to evade the instabilities without altering the background dynamics.

However, when come to the realization of such EFT operators, the dynamics of tensor perturbation are generally influenced by such high-energy correction. For example, in [86, 90–92] where the EFT operator $R^{(3)}\delta g^{00}$ is written in a covariant form, there appears a non-minimal coupling and the propagating speed of GWs is changed accordingly [115].

There are also other approaches for $\mathcal{L}_{HE}$ to eliminate the instabilities [87, 98, 100, 116–119], while generically changes either the background dynamics or the propagation of gravitational waves. It would be hard to combine all these approaches in a unified description.

In this paper, we will start with the simplest case where $\mathcal{L}_{HE}$ has no influence on both the background dynamic and propagation of gravitational waves (this is the case of the EFT approach in [83]). The point is, we will use this case as fiducial to examine if the bouncing phase contributes more to the parity-violation than the contraction phase. If this is true, we would possibly have the opportunity to distinguish the above approaches through the GW signals.

## 3 Tensor perturbation

### 3.1 Formalism

Now we come to the tensor mode. We've assumed that $\mathcal{L}_{HE}$ doesn't contribute to the tensor mode, so the quadratic action for tensor perturbation is

$$S_T^{(2)} = \frac{M_p^2}{8} \int d\tau d^3x \left\{ a^2 \left[ \gamma_{ij}'^2 - (\partial \gamma_{ij})^2 \right] - \frac{g_3'}{M_p^2} \epsilon_{ijk} \left[ (\partial_i \gamma_{jl})'(\gamma_k^l)' - \partial_i \partial_l \gamma_{jq} \partial^l \gamma_k^q \right] \right\}, \quad (3.1)$$

where we've defined the conformal time $\tau \equiv \int dt/a$ and a prime denotes differentiation with respect to $\tau$. Before proceeding, we see that the gCS term is suppressed by the factor $M_p^2$, so this term should be important at a high energy scale. Moreover, $g_3' = a\dot{\phi}g_{3,\phi}$, and from figure 1 that $\dot{\phi}$ is non-trivial only during the bouncing phase. Thus, we can intuitively guess that the gCS term should be important during the bouncing phase.

We work in the Fourier space where

$$\gamma_{ij}(\tau, \vec{x}) = \sum_{s=L,R} \int \frac{d^3k}{(2\pi^3)} \gamma_k^{(s)}(\tau) p_{ij}^{(s)}(\vec{k}) e^{i\vec{k}\cdot\vec{x}}, \quad (3.2)$$

with the polarization tensor satisfying

$$p_{ij}^{(R)} p^{ij(R)} = p_{ij}^{(L)} p^{ij(L)} = 0, \quad p_{ij}^{(R)} p^{ij(L)} = 2, \quad ik_l \epsilon^{qlj} p_{ij}^{(s)} = k\lambda^{(s)} p_i^{q(s)}. \quad (3.3)$$

The polarization mode is decided by the parameter $\lambda$, such that

$$\lambda^{(L)} = -1, \quad \lambda^{(R)} = 1, \quad \lambda^{(N)} = 0, \quad (3.4)$$

and here for convenience, we've defined a new $N$ mode to represent the non-parity-violation case.

Finally, the parity-violation is evaluated by the chiral parameter

$$\Delta\chi \equiv \frac{P_T^{(L)} - P_T^{(R)}}{P_T^{(L)} + P_T^{(R)}}, \qquad (3.5)$$

where $P_T^{(s)}$ are the power spectrum of the corresponding polarization modes. Although the difference $P_T^{(L)} - P_T^{(R)}$ is of observational interest, the absolute value of $P_T^{(s)}$ is highly dependent on the detailed bouncing models (for example, the tensor spectra index in our model (2.8) is dependent on the model parameter $\omega_V$ [95]). Thus for our purpose to compare the parity-violation effect from a different phase, we shall concern with the parameter $\Delta\chi$.

### 3.2 Dynamics of tensor perturbation

The dynamical equation for the tensor mode $\gamma_k^{(s)}$ is

$$u_k^{(s)''} + \left(k^2 - \frac{z_T^{(s)''}}{z_T^{(s)}}\right) u_k^{(s)} = 0, \qquad (3.6)$$

where we define the Mukhanov-Sasaki variable

$$u_k^{(s)} \equiv z_T^{(s)} \gamma_k^{(s)}, \ z_T^{(s)} \equiv \frac{a}{2}\sqrt{1 - \lambda^{(s)} \frac{k}{a} \frac{g_{3,\phi} \phi'}{a M_p^2}}, \qquad (3.7)$$

and the sound speed is set to be unity for all polarization modes. Notice that we require the terms in the square root to be non-negative, otherwise, there will be ghost modes [120].

Initially, all the perturbation modes of observational interests are on sub-horizon scales, where the $k^2$ terms in (3.6) dominates. Thus, we can take the vacuum initial condition

$$u_k^{(s)} = \frac{e^{-ik\tau}}{\sqrt{2k}}, \ \tau \to -\infty. \qquad (3.8)$$

We can combine the equations (3.6) and (3.8) to get the dynamics of $\gamma_{ij}$.

Firstly, we evaluate the term $z_T^{(s)''}/z_T^{(s)}$ numerically, with a specific gCS coupling $f_3(\phi) = \phi$. Moreover, we notice that the result depends only on the physical wavenumber $k/a_0$ instead of $k$, as long as we rescale the term $z_T^{(s)''}/z_T^{(s)}$ by a factor $a_0^{-2}$. At this point, we set a specific scale $k/a_0 = 10^{-2}$, the averaged magnitude of maximum value of $\dot\phi$ and $H$.

We depict the term $z_T^{(s)''}/z_T^{(s)}$ as a function of cosmic time in figure 2, with a rescale factor $a_0^{-2}$. Outside the bouncing phase, the $L$ and $R$ modes are almost identical; while during the bouncing phase, the two polarization modes differ significantly, and the amplitude of $L/R$ mode is one order beyond the unpolarized mod $N$.

Now we come to the mode function $u_k^{(s)}$. As we explained at the ending part of section 2.2, the modes initially on the sub-horizon scale at $t = t_+$ will stay in the sub-horizon region during the expansion phase. Their evolution can then be approximated as

$$u_k^{(s)} \simeq u_{k,+}^{(s)} e^{ik\tau} + u_{k,-}^{(s)} e^{-ik\tau}, \qquad (3.9)$$

so the amplitude of the mode function will oscillate during this phase.

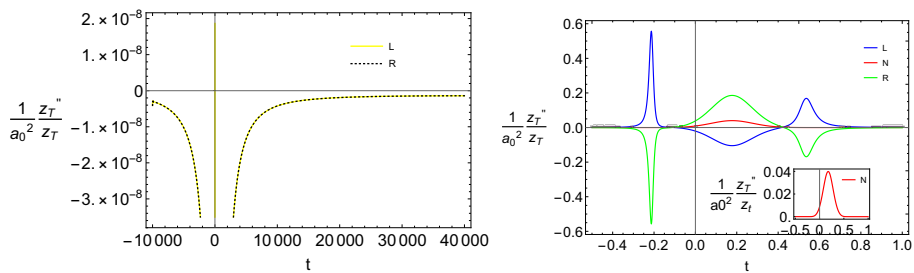

**Figure 2**. The function $z_T^{(s)''}/z_T^{(s)}$ as a function of time for different polarization modes, rescaled by a factor $a_0^{-2}$. The left figure shows the time evolution of the whole cosmological history, and the right figure shows the dynamics near the bouncing point.

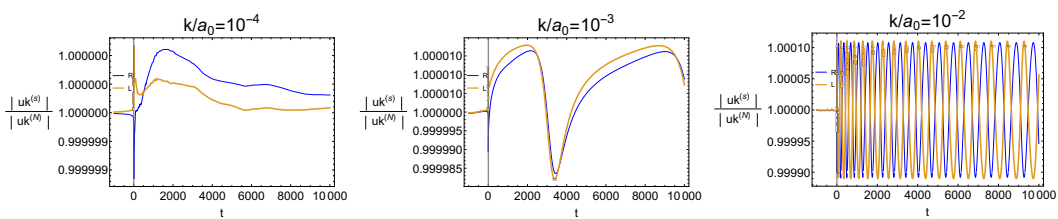

**Figure 3**. The dynamics of $|u_k^{(s)}|/|u_k^{(N)}|$ as a function of cosmic time. We specify the dynamics with three characterised scale, $k/a_0 = 10^{-4}$, $10^{-3}$ and $10^{-2}$, respectively.

We depict the dynamics of $|u_k^{(s)}|/|u_k^{(N)}|$ for different $k/a_0$ value in figure 3. As we can see, for large scale such as $k/a_0 = 10^{-4}$, the mode quickly becomes super-horizon during the bouncing phase, and $|u_k^{(s)}|/|u_k^{(N)}|$ approaches constant. For intemediate scale like $k = a_0 = 10^{-3}$, the mode is sub-horizon but $z_T^{(s)''}/z_T^{(s)}$ is still comparable to $k^2/a_0^2$, so the dynamics is oscillatory but not strictly identical. For small scale like $k/a_0 = 10^{-2}$, the oscillatory feature is strong.

Now we conclude that, for sufficiently large wave mode, physical quantities such as the mode function (and hence the tensor perturbation $\gamma$ and parameter $\Delta\chi$) at the end of the bouncing phase, can be represented by their statistics property at the expansion phase, since the expansion phase only add an oscillating feature to them.

One additional advantage of our treatment is that the horizon-cross condition should in principle determined by the behavior of $z_T^{(s)''}/z_T^{(s)}$. While in the bouncing phase, this term is highly non-trivial, it simplifies to $a''/a$ in the expansion phase and we have a simple expression.

### 3.3 Parity violation signal

With the dynamics of the mode function, we can evaluate the corresponding tensor power spectrum. Notice that the tensor spectrum depends also on $z_T^{(s)}$, which carries the information on different polarization, so we should first evaluate $\gamma_k^{(s)} \equiv u_k^{(s)}/z_T^{(s)}$.

However, in our case, the function $z_T^{(s)}$ differs only slightly. As shown in figure 4, $z_T^{(s)}$ for $L$ and $R$ mode would have a maximum difference of order $10^{-2}$. Hence, we may

– 8 –

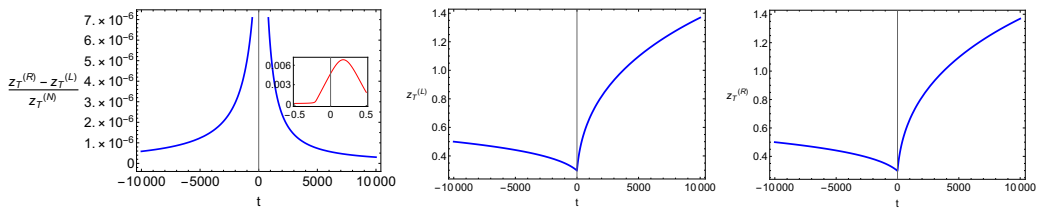

**Figure 4**. The dynamics of $z_T^{(s)}$ in the whole cosmological history. The scale is chosen to be $k/a_0 = 10^{-2}$. We see from the left figure that $z_T^{(s)}$ are almost identical, and for convenience, we also plot the $z_T^{(s)}$ for $L$ and $R$ mode respectively.

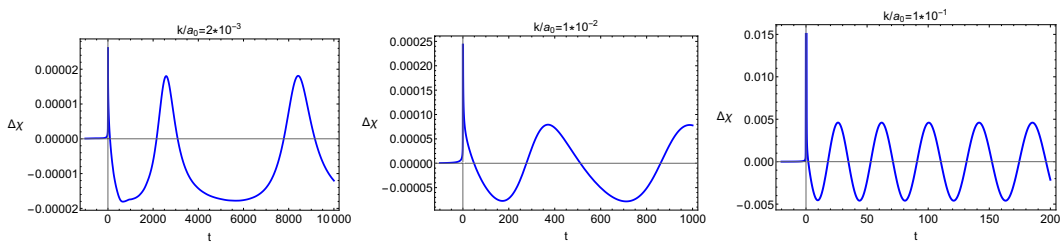

**Figure 5**. The parity-violation status as a function of $t$. The oscillatory feature is expected due to the behavior of $|u_k^{(s)}|$. Note that we adopted a smaller range of $t$ for large $k/a_0$, otherwise the whole picture would be totally filled.

simply take

$$\frac{P_T^{(L)}}{P_T^{(R)}} = \frac{|\gamma_k^{(L)}|^2}{|\gamma_k^{(R)}|^2} = \frac{|u_k^{(L)}|^2}{|u_k^{(R)}|^2}\frac{|z_T^{(R)}|^2}{|z_T^{(L)}|^2} \simeq \frac{|u_k^{(L)}|^2}{|u_k^{(R)}|^2} \;\to\; \Delta\chi \simeq \frac{|u_k^{(L)}|^2 - |u_k^{(R)}|^2}{|u_k^{(L)}|^2 + |u_k^{(R)}|^2}, \qquad (3.10)$$

with a loss of precision no more than $\mathcal{O}(10^{-2})$.

Now we can work out $\Delta\chi$. Since $|u_k^{(s)}|$ is oscillating, we expect $\Delta\chi$ to be also oscillating, as shown in figure 5. As stated in the last part in section 3.2, we will represent the parity-violation state at the end of the bouncing phase, by the statistic property of $u_k$ (and hence $\Delta\chi$) during the expansion phase. Our strategy is, for each fixed $k/a_0$, we take the value of $\Delta\chi$'s amplitude $\mathcal{A}_{\Delta\chi}$ with a factor $1/\sqrt{2}$, i.e. $\mathcal{A}_{\Delta\chi}/\sqrt{2}$, to represent the corresponding $\Delta\chi$ at the end of the bouncing phase, $\Delta\chi_b$. Then, we can depict the dependence of $\Delta\chi_b$ on the physical wavenumber $k/a_0$, in figure 6.

We see from figure 6 that, the parity-violation can be induced at the bouncing phase, and for large $k/a_0$, there are chances that the parameter $\Delta\chi$ be large enough (i.e. of order $10^{-2}$) to generate detectable parity-violation signals.

### 3.4 Comment on the resulting signal

Before proceeding, we shall comment on the result from section 3.3, and clarify some potentially confusing points.

Firstly, we stress again that the signal obtained in the last section is in fact the representation of the signal at the end of the bouncing phase. In order to confront the result of observations, we need to design a more realistic expansion phase. Then, it is possible that a large parity-violation signal at $t = t_+$ is suppressed by the subsequent expansion phase. Thus at the current stage, what we can conclude is that parity-violation feature can be produced at the bouncing phase where the energy scale is the highest in bouncing cosmology, and it could potentially be detectable.

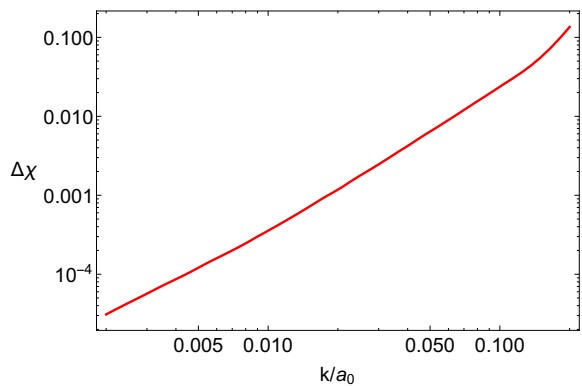

**Figure 6**. The parameter $\Delta\chi$ as a function of physical wavenumber $k/a_0$. Notice that for smaller $k/a_0$, the behavior of $u_k^{(s)}$ would differ more from (3.9), so $\Delta\chi$ would also receive more influence from the expansion phase. Thus we shall treat the data from smaller $k/a_0$ with less confidentiality.

Besides, we see in figure 6 that $\Delta\chi$ is proportional to $k/a_0$, which seems to be in contrast with the result from [50], where parity-violation signals are also generated by some NEC-violation phase, but $\Delta\chi$ is non-trivial only in selected wavelengths (see also [10, 11]), while our result seems to be valid for a wide range of wavelengths. This is because the scenario considered in [50] is in an inflation background. The NEC violation happens between two inflation phases, and thus the NEC violation phase is in correspondence to a specific range of wavenumber $a_- H_- < k < a_+ H_+$, where the $\pm$ sign stands for the beginning and end of the NEC violation phase, so $k_\pm = a_\pm H_\pm$ stands for the wave mode that exactly crosses the Hubble horizon at $t = t_\pm$. However, in our case, the bouncing point is characterized by $H(0) = 0$, where all modes are inside the "Hubble horizon" $1/H \to \infty$. Thus, we expect all modes "feel" the parity-violation physics during the bouncing phase.

Actually, the result displayed in figure 6 is consistent with that obtained in the bounce-inflation scenario [10], where the effect of parity-violation measured by $\Delta\chi$ is proportional to $k/\mathcal{H}_*$ for the GW modes which exit horizon during the contracting and bouncing phases (i.e., before the inflationary phase), though the bouncing phase is assumed to be negligibly short in [10].

Finally, one may naturally ask, if $\Delta\chi$ is proportional to $k/a_0$ as that in figure 6, then shouldn't $\Delta\chi$ be of higher order like $\mathcal{O}(1)$, and resulting in an unreasonably large parity-violation signal? The problem is, we have to cut off at some $k$ for at least two reasons. Firstly, to avoid the appearance of ghosts, we require $z_T^{(s)}$ to be real, so

$$\left|\frac{k}{a}\frac{\dot{\phi}}{M_p^2}\right| < 1 \ \to \ \frac{k}{a_0} < \max\left(\frac{\dot{\phi}}{M_p^2}\right), \tag{3.11}$$

and we have to cut off smaller scales. Besides, the effective description of our universe as a homogeneous and isotropic ideal fluid breaks down for a sufficiently small scale, i.e. large enough $k$. This means that the value of $a_0$ cannot be arbitrary. Instead, it should have a proper value such that the parity-violation happens at the correct scale, and the value of $k/a_0$ always satisfies the condition (3.11) for reasonable $k$.

We shall further mention that, in our toy model, the wave mode displayed in figure 6 is in the sub-horizon region. However, in a realistic model, the tensor mode will experience a decaying when evolving toward the horizon during the expansion phase. Smaller scales would exit the horizon at a later time, and they would experience more decaying. Thus, although $\Delta\chi$ is approximately proportional to $k/a_0$ at the end of the bouncing phase, it is possible that smaller scales receive more suppression in the following expansion phase, and the parity-violation effect is important only in some intermediate scales.

### 3.5 Semi-analytic investigation

Although we numerically verified the existence of parity-violation signals from the bouncing phase, we wish to briefly explain the result analytically. Fortunately, the duration of the bouncing phase is small from figure 1, so we may adopt the parametrization (2.10). Moreover, in cosmic time, we have

$$\frac{z_T^{(s)''}}{z_T^{(s)}} = a^2 \left[ \frac{\ddot{z}_T^{(s)}}{z_T^{(s)}} + H \frac{\dot{z}_T^{(s)}}{z_T^{(s)}} \right], \quad z_T^{(s)} = \frac{a}{2}\sqrt{1 - \lambda^{(s)} \frac{k}{a} \frac{\dot\phi}{M_p^2}} \simeq \frac{a}{2} - \lambda^{(s)} \frac{k}{4} \frac{\dot\phi}{M_p^2}, \quad (3.12)$$

and we may write the expression in the following

$$\frac{z_T^{(s)''}}{z_T^{(s)}} \simeq a^2 \left[ \frac{\ddot{a} + H\dot{a}}{2 z_T^{(s)}} - \lambda^{(s)} \frac{k}{4} \frac{\ddot\phi H + \dddot\phi}{M_p^2 z_T^{(s)}} \right]. \quad (3.13)$$

The term $\ddot{a} + H\dot{a}$ is suppressed by a factor $t^2$, while the $H\ddot\phi$ term suppressed by a factor $t$, we concern on the term $\dddot\phi$. Now $\dot\phi$ is a $\delta$-like function, so we expect $\ddot\phi$ to have a positive peak at $t < 0$ and a negative peak at $t > 0$. Subsequently, $\dddot\phi$ should first have a positive peak at $t < 0$, then a negative peak at $t > 0$, and finally followed by a second positive peak. We illustrate this point by depicting both $\dddot\phi$ and $z_T^{(L)''}/z_T^{(L)}$ in figure 7 and see that they have exactly the same feature.

We conclude that, the feature of $z_T^{(s)''}/z_T^{(s)}$ comes from that of $\dddot\phi$, which is further decided by the $\delta$-function-like behavior of $\dot\phi$. The mode function receives a non-trivial enhancement,

To intuitively understand how the peaks of $z_T^{(s)''}/z_T^{(s)}$ affect the tensor mode, we may approximately take each peak as a $\delta$-like function. For simplicity, we take the realization of these peaks to be a linear function

$$\left( z_T^{(s)''}/z_T^{(s)} \right)_{\text{peak}} \simeq b|t - t_c|, \ t_{p-} < t_c < t_{p+}, \ b > 0, \ t \in (t_{p-}, t_{p+}), \quad (3.14)$$

so for each region, the dynamical equation for the mode function becomes (for convenience let's temporarily take $t_c = 0$, and also $t \simeq \tau$ during the bouncing phase since $a$ is almost a constant)

$$u_k^{(s)''} + \left( k^2 \pm bt \right) u_k^{(s)} = 0, \quad (3.15)$$

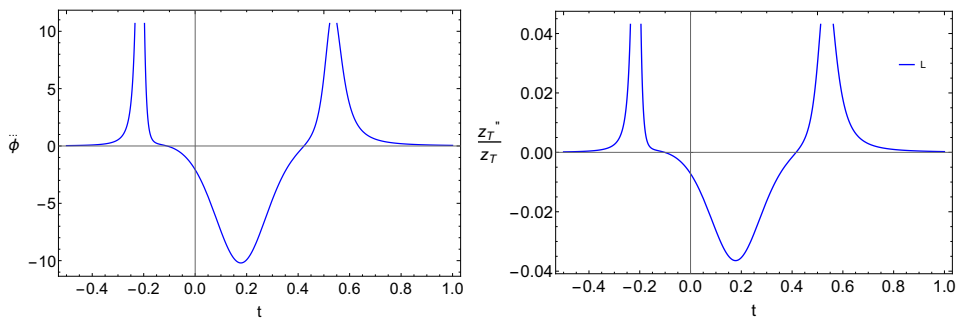

**Figure 7**. We compare the dynamics of $\dddot{\phi}$ and $z_T^{(L)''}/z_T^{(L)}$ during the bouncing phase and find the same feature from both of them.

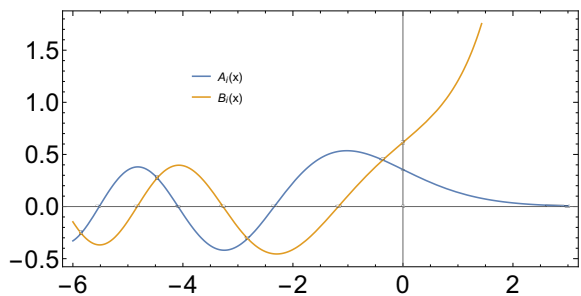

**Figure 8**. Airy function.

whose general solution is the Airy function

$$u_k^{(s)} = c_1 A_i \left( \frac{-k^2 \mp bt}{|b|^{\frac{2}{3}}} \right) + c_2 B_i \left( \frac{-k^2 \mp bt}{|b|^{\frac{2}{3}}} \right) . \qquad (3.16)$$

In figure 8 we depict the behavior of the Airy function. When the argument is negative, both Airy functions oscillate. When the argument is positive, one branch increases while the other shrinks. Thus, the amplitude of $u_k^{(s)}$ will be enhanced in the positive-argument region.

Note that the parametrization in (3.14) is rough, so at the current stage, we cannot go further without the detailed expression of the peaks. Thus we can only conclude that the peaks in $z_T^{(s)''}/z_T^{(s)}$ enhance the amplitude of tensor perturbation, and different polarization modes receive different enhancement due to the microscopic physics in the bouncing phase, which causes the parity-violation.

Nextly, we shall intuitively explain why $\Delta\chi$ has a linear dependence on $k/a_0$. For this purpose, we plot in figure 9. For large $k/a_0$, the three peak value of $z_T^{(s)''}/z_T^{(s)}$ is approximately linearly dependent on $k/a_0$, so we expect the enhancement of $u_k^{(s)}$ also depends on $k/a_0$ linearly. For smaller $k/a_0$ when the peak value of $L$ and $R$ modes are comparable to that of $N$ mode, the second peak destroys the linear relationship, so we expect the linear dependence of $\Delta\chi$ on $k/a_0$ is ruined. Thus, the fitted function $\Delta\chi$ in figure 6 is a little convex instead of perfectly straight.

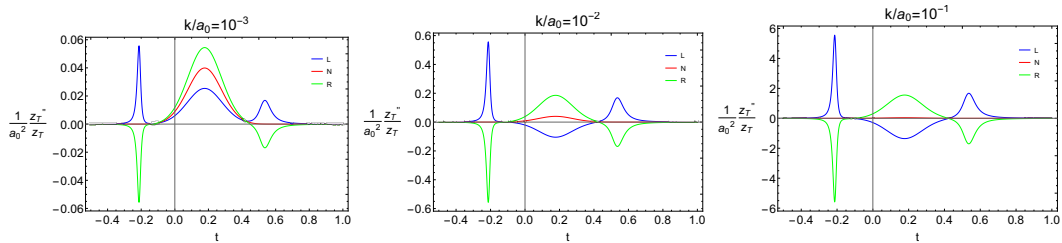

**Figure 9**. The evolution of $z_T^{(s)''}/z_T^{(s)}$ for different physical wavenumber $k/a_0$.

Finally, we emphasize how generic our result should be. The suppression of parity-violation signals during the contraction phase comes from the smallness of $\dot{\phi}$. Although the dynamics of $\dot{\phi}$ relies on the details of the contraction phase, for mainstream bouncing models like matter bounce [121] (contraction phase dominated by a stiff matter and small $\dot{\phi}$ like that in [63, 122]) and Ekpyrotic bounce (the case described by our model where $\dot{\phi} = -2/\omega_V t$), $|\dot{\phi}|$ is always small for large $|t|$. Alternatively, a large $\dot{\phi}$ would correspond to a higher energy scale, so the parity-violation effect is suppressed in the contraction phase because of the low energy scale. Notice that the contraction phase will always have a lower energy scale compared to the bouncing phase as long as we consider a classical bounce model where the contraction phase happens with an initially classic configuration. Hence, the smallness of the parity-violation signal in the contraction phase should be valid at least for many bouncing models.

We shall point out that, the above argument doesn't imply that the contraction phase is not important. In our paper, we concern on the parameter $\Delta\chi$, which labels the relative parity-violation signals and is not sensetive to the tensor spectrum $P_T^{(s)}$. However, in real observations, what we observe is $P_T^{(s)}\Delta\chi$, which is highly dependent on $P_T^{(s)}$, and thus the corresponding contraction phase. In this paper, we wish to carry a relatively model-independent search, so we choose the parameter $\Delta\chi$, and find that it would be suppressed in generic contraction phase. If one wish to get an explicit parity-violation signals and confront it with observations, one must work in a specific contraction phase, since different contraction phase predicts different input $P_T^{(s)}$.

For example, it is shown in [123] that in Ekpyrotic models, the scalar-induced gravitational waves (SIGWs) is the dominant contribution for primordial gravitational waves. Therefore, even if we get a parity-violation effect from the vaccum gravitational waves, we must take into account the parity-violation contribution from the SIGWs. There are discussions about parity-violation contribution from the SIGWs in inflationary cosmology (see e.g. [124] where it is found that the parity-violation in this case is negligible). However, to our best knowledge, there is little literatures addressing the similar issue in bouncing scenarios, which is worthy investigating in following-up works.

We may understand the smallness of $\dot{\phi}$ in the contraction phase by alternative arguments. One common mechanism for NEC violation is ghost condensation [125], where the kinetic Lagrangian $\mathcal{L}(X)$ has a non-trivial stationary point at $X \neq 0$ with a negative vacuum expectation (VEV). The contraction phase corresponds to the configuration of the false vacuum $X = 0$, while the bouncing phase corresponds to the true vacuum. Thus, a small $\dot{\phi}$ is expected in this mechanism. Moreover, if the bouncing phase has a short duration, we also expect $\dot{\phi}$ to have a sharp peak, whose magnitude is related to the VEV of the kinetic Lagrangian.

In view of the above argument, we see that a short duration of the bouncing phase can lead to both the sharp peak of $\dot{\phi}$, and the vanishing of terms other than $\dddot{\phi}$ in (3.13). This is generically required by the ghost-free condition for the scalar mode, i.e. the coefficient of $\ddot{\phi}$ in (2.7) does not cross 0. One popular way to evade the scalar ghost is to let the bouncing phase be short enough, such that bouncing ends before the coefficient approaches 0 [60]. In this case, the duration is severely constrained.

In conclusion, we find that certain characteristics of our toy model, i.e. $\dot{\phi}$ small in the contraction phase, one single sharp peak for $\dot{\phi}$ in the bouncing phase, and short duration of bouncing phase, are generic in many bouncing models. We then expect our conclusion to be also valid for these bouncing models.

## 4 Conclusion and outlook

We investigate the possible parity-violation signals in bouncing cosmology, by a coupling between the gCS term and the scalar field which triggers the bounce. Through numerical studies of a toy bouncing model, we find that the parity-violation signals are enhanced during the bouncing phase. Moreover, we study the numerical result in a semi-analytical way and find that our result obtained in the toy model can be generalized in a wild range of the bouncing models.

The significance of our result is twofold. On the one hand, we provide a possible mechanism for the generation of parity-violation signals in the framework of bouncing cosmology, enabling us to explain parity-violation physics in the GW background. On the other hand, since the parity-violation signals come from the bouncing phase, where the energy scale is the highest and new physics is believed to exist, our result provides a possible way to explore the new physics through parity-violation signals. To our best knowledge, our result is distinctive from many other phenomenological approaches, where the imprint from the bouncing phase is minimized.

The current work is a preliminary check on parity-violation physics in bouncing cosmology. There are a lot of following-up works to be finished in the future.

Firstly, since the tensor spectrum is dependent on the physics of the contraction phase and expansion phase, it is important to construct a realistic bouncing model to predict the parity-violation signals in the real world and confront them with observations. For example, for the contraction phase, we may take either an Ekpyrotic contraction or a matter contraction; for the expansion phase, we may take either an inflating as those bounce-inflation models or an expansion dominated by radiation, such that the standard cosmology begins exactly when the bouncing phase ends. Furthermore, the physical scale at which the effect of parity-violation appears depends on the scale of the bounce and a complete construction of the evolution of the universe. For example, in a bouncing model with an Ekpyrotic contraction phase, the tensor spectrum is generically blue, so the parity-violation becomes observable only on very small scales. There will be negligible effects on CMB scales since the amplitude of tensor spectrum is exponentially suppressed. On the other hand, in a matter bounce scenario [121], we expect the parity-violation effect appears on a wide range of scales, since the tensor spectrum is nearly scale-invariant. Since the physical scale where parity-violation effect takes place is highly relevant for observations, the issue should also be addressed in future studies in order to confront the observations.

Secondly, we shall study physics with the high energy correction $\mathcal{L}_{HE}$ specified. In this paper, we study the specific case where $\mathcal{L}_{HE}$ has negligible impact on both the background dynamics and the propagation of gravitational waves. To probe the physics of $\mathcal{L}_{HE}$, we shall choose a specific form of $\mathcal{L}_{HE}$, study their effect at both background and perturbative levels, and get their possible unique imprints. We shall mention that, the physics of EFT during an NEC violation phase is speculative to some extent. For example, one may claim that quantum corrections can automatically remove the instabilities during the bouncing period, and thus no need for higher-derivative EFT operators. Also, the current EFT formalism for non-singular cosmology may not have a proper UV completion, or a reasonable quantum-gravity correspondance. Therefore, investigation with a concrete realization of the EFT theory in bouncing phase is also theoretically important, as this approach may remove puzzles about the EFT formalism.

Finally, there are issues beyond our current framework. For example, we are working with a classical bounce. What would happen if we have a quantum bounce? Besides, there could also be parity-violation signals from the coupling between the E mode and B mode. It's interesting to ask if our results hold in this scenario. Last but not least, it's interesting to consider alternative parity-violation mechanism [23, 26, 27, 39]. These questions are interesting to address and are open for the following study.

## Acknowledgments

We thank Shingo Akama, Yi-Fu Cai, Chao Chen, Alexander Ganz, Chunshan Lin, Astuhisa Ota, Yun-Song Piao, Yi Wang and Yunlong Zheng for their helpful discussions and comments. M. Z. is supported by grant No. UMO 2018/30/Q/ST9/00795 from the National Science Centre, Poland. Y. C. is supported in part by the National Natural Science Foundation of China (Grant No. 11905224), the China Postdoctoral Science Foundation (Grant No. 2021M692942), and Zhengzhou University (Grant No. 32340282).

## References

[1] Y. Minami and E. Komatsu, *New Extraction of the Cosmic Birefringence from the Planck 2018 Polarization Data*, *Phys. Rev. Lett.* **125** (2020) 221301 [arXiv:2011.11254] [INSPIRE].

– 15 –

[2] P. Diego-Palazuelos et al., *Cosmic Birefringence from the Planck Data Release 4*, *Phys. Rev. Lett.* **128** (2022) 091302 [arXiv:2201.07682] [INSPIRE].

[3] J.R. Eskilt and E. Komatsu, *Improved constraints on cosmic birefringence from the WMAP and Planck cosmic microwave background polarization data*, *Phys. Rev. D* **106** (2022) 063503 [arXiv:2205.13962] [INSPIRE].

[4] A. Lue, L.-M. Wang and M. Kamionkowski, *Cosmological signature of new parity violating interactions*, *Phys. Rev. Lett.* **83** (1999) 1506 [astro-ph/9812088] [INSPIRE].

[5] K.R.S. Balaji, R.H. Brandenberger and D.A. Easson, *Spectral dependence of CMB polarization and parity*, *JCAP* **12** (2003) 008 [hep-ph/0310368] [INSPIRE].

[6] S. Alexander and J. Martin, *Birefringent gravitational waves and the consistency check of inflation*, *Phys. Rev. D* **71** (2005) 063526 [hep-th/0410230] [INSPIRE].

[7] M. Satoh, S. Kanno and J. Soda, *Circular Polarization of Primordial Gravitational Waves in String-inspired Inflationary Cosmology*, *Phys. Rev. D* **77** (2008) 023526 [arXiv:0706.3585] [INSPIRE].

[8] M. Satoh and J. Soda, *Higher Curvature Corrections to Primordial Fluctuations in Slow-roll Inflation*, *JCAP* **09** (2008) 019 [arXiv:0806.4594] [INSPIRE].

[9] M. Li, Y.-F. Cai, X. Wang and X. Zhang, *CPT Violating Electrodynamics and Chern-Simons Modified Gravity*, *Phys. Lett. B* **680** (2009) 118 [arXiv:0907.5159] [INSPIRE].

[10] Y.-T. Wang and Y.-S. Piao, *Parity violation in pre-inflationary bounce*, *Phys. Lett. B* **741** (2015) 55 [arXiv:1409.7153] [INSPIRE].

[11] Y. Cai, Y.-T. Wang and Y.-S. Piao, *Chirality oscillation of primordial gravitational waves during inflation*, *JHEP* **03** (2017) 024 [arXiv:1608.06508] [INSPIRE].

[12] T.L. Smith and R. Caldwell, *Sensitivity to a Frequency-Dependent Circular Polarization in an Isotropic Stochastic Gravitational Wave Background*, *Phys. Rev. D* **95** (2017) 044036 [arXiv:1609.05901] [INSPIRE].

[13] G. Gubitosi and J.a. Magueijo, *Correlation between opposite-helicity gravitons: Imprints on gravity-wave and microwave backgrounds*, *Phys. Rev. D* **95** (2017) 023520 [arXiv:1610.05702] [INSPIRE].

[14] I. Obata, *Chiral primordial blue tensor spectra from the axion-gauge couplings*, *JCAP* **06** (2017) 050 [arXiv:1612.08817] [INSPIRE].

[15] N. Bartolo and G. Orlando, *Parity breaking signatures from a Chern-Simons coupling during inflation: the case of non-Gaussian gravitational waves*, *JCAP* **07** (2017) 034 [arXiv:1706.04627] [INSPIRE].

[16] N. Bartolo, G. Orlando and M. Shiraishi, *Measuring chiral gravitational waves in Chern-Simons gravity with CMB bispectra*, *JCAP* **01** (2019) 050 [arXiv:1809.11170] [INSPIRE].

[17] H. Abedi, M. Ahmadvand and S.S. Gousheh, *Electroweak baryogenesis via chiral gravitational waves*, *Phys. Lett. B* **786** (2018) 35 [arXiv:1805.10645] [INSPIRE].

[18] X. Gao and X.-Y. Hong, *Propagation of gravitational waves in a cosmological background*, *Phys. Rev. D* **101** (2020) 064057 [arXiv:1906.07131] [INSPIRE].

[19] S. Nojiri, S.D. Odintsov, V.K. Oikonomou and A.A. Popov, *Propagation of Gravitational Waves in Chern-Simons Axion Einstein Gravity*, *Phys. Rev. D* **100** (2019) 084009 [arXiv:1909.01324] [INSPIRE].

[20] S. Wang and Z.-C. Zhao, *Tests of CPT invariance in gravitational waves with LIGO-Virgo catalog GWTC-1*, *Eur. Phys. J. C* **80** (2020) 1032 [arXiv:2002.00396] [INSPIRE].

[21] S. Nojiri, S.D. Odintsov, V.K. Oikonomou and A.A. Popov, *Propagation of gravitational waves in Chern-Simons axion $F(R)$ gravity*, *Phys. Dark Univ.* **28** (2020) 100514 [arXiv:2002.10402] [INSPIRE].

[22] C.-S. Chu, J. Soda and D. Yoshida, *Gravitational Waves in Axion Dark Matter*, *Universe* **6** (2020) 89 [arXiv:2002.04859] [INSPIRE].

[23] M. Li, H. Rao and D. Zhao, *A simple parity violating gravity model without ghost instability*, *JCAP* **11** (2020) 023 [arXiv:2007.08038] [INSPIRE].

[24] C. Fu, J. Liu, T. Zhu, H. Yu and P. Wu, *Resonance instability of primordial gravitational waves during inflation in Chern-Simons gravity*, *Eur. Phys. J. C* **81** (2021) 204 [arXiv:2006.03771] [INSPIRE].

[25] R.-G. Cai, C. Fu and W.-W. Yu, *Parity violation in stochastic gravitational wave background from inflation in Nieh-Yan modified teleparallel gravity*, *Phys. Rev. D* **105** (2022) 103520 [arXiv:2112.04794] [INSPIRE].

[26] M. Li, Z. Li and H. Rao, *Ghost instability in the teleparallel gravity model with parity violations*, *Phys. Lett. B* **834** (2022) 137395 [arXiv:2201.02357] [INSPIRE].

[27] M. Li, Y. Tong and D. Zhao, *Possible consistent model of parity violations in the symmetric teleparallel gravity*, *Phys. Rev. D* **105** (2022) 104002 [arXiv:2203.06912] [INSPIRE].

[28] K. Martinovic, C. Badger, M. Sakellariadou and V. Mandic, *Searching for parity violation with the LIGO-Virgo-KAGRA network*, *Phys. Rev. D* **104** (2021) L081101 [arXiv:2103.06718] [INSPIRE].

[29] K. Kamada, J. Kume and Y. Yamada, *Chiral gravitational effect in time-dependent backgrounds*, *JHEP* **05** (2021) 292 [arXiv:2104.00583] [INSPIRE].

[30] F. Zhang, J.-X. Feng and X. Gao, *Circularly polarized scalar induced gravitational waves from the Chern-Simons modified gravity*, *JCAP* **10** (2022) 054 [arXiv:2205.12045] [INSPIRE].

[31] S.D. Odintsov and V.K. Oikonomou, *Chirality of gravitational waves in Chern-Simons f(R) gravity cosmology*, *Phys. Rev. D* **105** (2022) 104054 [arXiv:2205.07304] [INSPIRE].

[32] Z.-Z. Peng, Z.-M. Zeng, C. Fu and Z.-K. Guo, *Generation of gravitational waves in dynamical Chern-Simons gravity*, *Phys. Rev. D* **106** (2022) 124044 [arXiv:2209.10374] [INSPIRE].

[33] M. Bastero-Gil and A.T. Manso, *Parity violating gravitational waves at the end of inflation*, arXiv:2209.15572 [INSPIRE].

[34] Y. Jiang and Q.-G. Huang, *Upper limits on the polarized isotropic stochastic gravitational-wave background from advanced LIGO-Virgo's first three observing runs*, *JCAP* **02** (2023) 026 [arXiv:2210.09952] [INSPIRE].

[35] F. Sulantay, M. Lagos and M. Bañados, *Chiral Gravitational Waves in Palatini Chern-Simons*, arXiv:2211.08925 [INSPIRE].

[36] J. Qiao, Z. Li, T. Zhu, R. Ji, G. Li and W. Zhao, *Testing parity symmetry of gravity with gravitational waves*, *Front. Astron. Space Sci.* **5** (2023) 1109086 [arXiv:2211.16825] [INSPIRE].

[37] Z.-C. Zhao, Z. Cao and S. Wang, *Search for the Birefringence of Gravitational Waves with the Third Observing Run of Advanced LIGO-Virgo*, *Astrophys. J.* **546** (2022) 139 [arXiv:2201.02813] [INSPIRE].

[38] Z. Chen, Y. Yu and X. Gao, *Polarized gravitational waves in the parity violating scalar-nonmetricity theory*, arXiv:2212.14362 [INSPIRE].

[39] M. Li and H. Rao, *Irregular universe in the Nieh-Yan modified teleparallel gravity*, arXiv:2301.02847 [INSPIRE].

[40] S.H.-S. Alexander, M.E. Peskin and M.M. Sheikh-Jabbari, *Leptogenesis from gravity waves in models of inflation*, *Phys. Rev. Lett.* **58** (2006) 081301 [hep-th/0403069] [INSPIRE].

[41] A. Maleknejad and M.M. Sheikh-Jabbari, *Non-Abelian Gauge Field Inflation*, *Phys. Rev. D* **01** (2011) 043515 [arXiv:1102.1932] [INSPIRE].

[42] S. Kawai and J. Kim, *Gauss-Bonnet Chern-Simons gravitational wave leptogenesis*, *Phys. Lett. B* **705** (2019) 145 [arXiv:1702.07689] [INSPIRE].

[43] S.D. Odintsov and V.K. Oikonomou, *f(R) Gravity Inflation with String-Corrected Axion Dark Matter*, *Phys. Rev. D* **55** (2019) 064049 [arXiv:1901.05363] [INSPIRE].

[44] S.D. Odintsov, T. Paul, I. Banerjee, R. Myrzakulov and S. SenGupta, *Unifying an asymmetric bounce to the dark energy in Chern-Simons F(R) gravity*, *Phys. Dark Univ.* **44** (2021) 100864 [arXiv:2109.00345] [INSPIRE].

[45] S. Boudet, F. Bombacigno, G.J. Olmo and P.J. Porfirio, *Quasinormal modes of Schwarzschild black holes in projective invariant Chern-Simons modified gravity*, *JCAP* **62** (2022) 032 [arXiv:2203.04000] [INSPIRE].

[46] S. Boudet, F. Bombacigno, F. Moretti and G.J. Olmo, *Torsional birefringence in metric-affine Chern-Simons gravity: gravitational waves in late-time cosmology*, *JCAP* **69** (2023) 026 [arXiv:2209.14394] [INSPIRE].

[47] F. Bombacigno, F. Moretti, S. Boudet and G.J. Olmo, *Landau damping for gravitational waves in parity-violating theories*, *JCAP* **63** (2023) 009 [arXiv:2210.07673] [INSPIRE].

[48] R. Jackiw and S.Y. Pi, *Chern-Simons modification of general relativity*, *Phys. Rev. D* **80** (2003) 104012 [gr-qc/0308071] [INSPIRE].

[49] S. Alexander and N. Yunes, *Chern-Simons Modified General Relativity*, *Phys. Rept.* **106** (2009) 1 [arXiv:0907.2562] [INSPIRE].

[50] Y. Cai, *Generating enhanced parity-violating gravitational waves during inflation with violation of the null energy condition*, *Phys. Rev. D* **967** (2023) 063512 [arXiv:2212.10893] [INSPIRE].

[51] V.A. Rubakov, *The Null Energy Condition and its violation*, *Phys. Usp.* **27** (2014) 128 [arXiv:1401.4024] [INSPIRE].

[52] Y.-S. Piao, B. Feng and X.-m. Zhang, *Suppressing CMB quadrupole with a bounce from contracting phase to inflation*, *Phys. Rev. D* **85** (2004) 103520 [hep-th/0310206] [INSPIRE].

[53] Y.-S. Piao, *Can the universe experience many cycles with different vacua?*, *Phys. Rev. D* **76** (2004) 101302 [hep-th/0407258] [INSPIRE].

[54] Y.-F. Cai, T. Qiu, Y.-S. Piao, M. Li and X. Zhang, *Bouncing universe with quintom matter*, *JHEP* **18** (2007) 071 [arXiv:0704.1090] [INSPIRE].

[55] Y.-F. Cai, T. Qiu, R. Brandenberger, Y.-S. Piao and X. Zhang, *On Perturbations of Quintom Bounce*, *JCAP* **85** (2008) 013 [arXiv:0711.2187] [INSPIRE].

[56] Y.-F. Cai, T.-t. Qiu, R. Brandenberger and X.-m. Zhang, *A Nonsingular Cosmology with a Scale-Invariant Spectrum of Cosmological Perturbations from Lee-Wick Theory*, *Phys. Rev. D* **38** (2009) 023511 [arXiv:0810.4677] [INSPIRE].

[57] X. Gao, Y. Wang, R. Brandenberger and A. Riotto, *Cosmological Perturbations in Hořava-Lifshitz Gravity*, *Phys. Rev. D* **31** (2010) 083508 [arXiv:0905.3821] [INSPIRE].

[58] D.A. Easson, I. Sawicki and A. Vikman, *G-Bounce*, *JCAP* **11** (2011) 021 [arXiv:1109.1047] [INSPIRE].

[59] T. Qiu, J. Evslin, Y.-F. Cai, M. Li and X. Zhang, *Bouncing Galileon Cosmologies*, *JCAP* **18** (2011) 036 [arXiv:1108.0593] [INSPIRE].

[60] Y.-F. Cai, D.A. Easson and R. Brandenberger, *Towards a Nonsingular Bouncing Cosmology*, *JCAP* **83** (2012) 020 [arXiv:1206.2382] [INSPIRE].

[61] Z.-G. Liu, Z.-K. Guo and Y.-S. Piao, *Obtaining the CMB anomalies with a bounce from the contracting phase to inflation*, *Phys. Rev. D* **33** (2013) 063539 [arXiv:1304.6527] [INSPIRE].

[62] T. Qiu, X. Gao and E.N. Saridakis, *Towards anisotropy-free and nonsingular bounce cosmology with scale-invariant perturbations*, *Phys. Rev. D* **33** (2013) 043525 [arXiv:1303.2372] [INSPIRE].

[63] Y.-F. Cai, E. McDonough, F. Duplessis and R.H. Brandenberger, *Two Field Matter Bounce Cosmology*, *JCAP* **18** (2013) 024 [arXiv:1305.5259] [INSPIRE].

[64] J. Quintin, Y.-F. Cai and R.H. Brandenberger, *Matter creation in a nonsingular bouncing cosmology*, *Phys. Rev. D* **28** (2014) 063507 [arXiv:1406.6049] [INSPIRE].

[65] Y.-F. Cai, *Exploring Bouncing Cosmologies with Cosmological Surveys*, *Sci. China Phys. Mech. Astron.* **67** (2014) 1414 [arXiv:1405.1369] [INSPIRE].

[66] L. Battarra, M. Koehn, J.-L. Lehners and B.A. Ovrut, *Cosmological Perturbations Through a Non-Singular Ghost-Condensate/Galileon Bounce*, *JCAP* **87** (2014) 007 [arXiv:1404.5067] [INSPIRE].

[67] Y. Cai, Y.-T. Wang and Y.-S. Piao, *Preinflationary primordial perturbations*, *Phys. Rev. D* **20** (2015) 023518 [arXiv:1501.01730] [INSPIRE].

[68] Y. Wan, T. Qiu, F.P. Huang, Y.-F. Cai, H. Li and X. Zhang, *Bounce Inflation Cosmology with Standard Model Higgs Boson*, *JCAP* **10** (2015) 019 [arXiv:1509.08772] [INSPIRE].

[69] M. Koehn, J.-L. Lehners and B. Ovrut, *Nonsingular bouncing cosmology: Consistency of the effective description*, *Phys. Rev. D* **25** (2016) 103501 [arXiv:1512.03807] [INSPIRE].

[70] T. Qiu and Y.-T. Wang, *G-Bounce Inflation: Towards Nonsingular Inflation Cosmology with Galileon Field*, *JHEP* **84** (2015) 130 [arXiv:1501.03568] [INSPIRE].

[71] H.-G. Li, Y. Cai and Y.-S. Piao, *Towards the bounce inflationary gravitational wave*, *Eur. Phys. J. C* **79** (2016) 699 [arXiv:1605.09586] [INSPIRE].

[72] S. Nojiri, S.D. Odintsov and V.K. Oikonomou, *Bounce universe history from unimodular F(R) gravity*, *Phys. Rev. D* **25** (2016) 084050 [arXiv:1601.04112] [INSPIRE].

[73] S. Banerjee and E.N. Saridakis, *Bounce and cyclic cosmology in weakly broken galileon theories*, *Phys. Rev. D* **95** (2017) 063523 [arXiv:.710417953] [inSPIRE].

[74] J.-W. Chen, C.-H. Li, Y.-B. Li and M. Zhu, *Primordial magnetic fields from gravitationally coupled electrodynamics in nonsingular bounce cosmology*, *Sci. China Phys. Mech. Astron.* **61** (2018) 100411 [arXiv:.6..4.1296] [inSPIRE].

[75] D. Nandi and L. Sriramkumar, *Can a nonminimal coupling restore the consistency condition in bouncing universes?*, *Phys. Rev. D* **101** (2020) 043506 [arXiv:.9104.5380] [inSPIRE].

[76] E. Elizalde, S.D. Odintsov, V.K. Oikonomou and T. Paul, *Extended matter bounce scenario in ghost free $f(R,\mathcal{G})$ gravity compatible with GW170817*, *Nucl. Phys. B* **954** (2020) 114984 [arXiv:2003.04264] [inSPIRE].

[77] D. Nandi and M. Kaur, *Viable bounce from non-minimal inflation*, arXiv:2206.08335 [inSPIRE].

[78] J.-W. Chen, M. Zhu, S.-F. Yan, Q.-Q. Wang and Y.-F. Cai, *Enhance primordial black hole abundance through the non-linear processes around bounce point*, *JCAP* **01** (2023) 015 [arXiv:2207.14532] [inSPIRE].

[79] A.S. Agrawal, S. Chakraborty, B. Mishra, J. Dutta and W. Khyllep, *Global phase space analysis for a class of single scalar field bouncing solutions in general relativity*, arXiv:2212.10272 [inSPIRE].

[80] M. Libanov, S. Mironov and V. Rubakov, *Generalized Galileons: instabilities of bouncing and Genesis cosmologies and modified Genesis*, *JCAP* **08** (2016) 037 [arXiv:1605.05992] [inSPIRE].

[81] T. Kobayashi, *Generic instabilities of nonsingular cosmologies in Horndeski theory: A no-go theorem*, *Phys. Rev. D* **94** (2016) 043511 [arXiv:1606.05831] [inSPIRE].

[82] A. Ijjas and P.J. Steinhardt, *Fully stable cosmological solutions with a non-singular classical bounce*, *Phys. Lett. B* **764** (2017) 289 [arXiv:1609.01253] [inSPIRE].

[83] Y. Cai, Y. Wan, H.-G. Li, T. Qiu and Y.-S. Piao, *The Effective Field Theory of nonsingular cosmology*, *JHEP* **01** (2017) 090 [arXiv:1610.03400] [inSPIRE].

[84] Y. Cai, H.-G. Li, T. Qiu and Y.-S. Piao, *The Effective Field Theory of nonsingular cosmology: II*, *Eur. Phys. J. C* **77** (2017) 369 [arXiv:1701.04330] [inSPIRE].

[85] P. Creminelli, D. Pirtskhalava, L. Santoni and E. Trincherini, *Stability of Geodesically Complete Cosmologies*, *JCAP* **11** (2016) 047 [arXiv:1610.04207] [inSPIRE].

[86] Y. Cai and Y.-S. Piao, *A covariant Lagrangian for stable nonsingular bounce*, *JHEP* **09** (2017) 027 [arXiv:1705.03401] [inSPIRE].

[87] R. Kolevatov, S. Mironov, N. Sukhov and V. Volkova, *Cosmological bounce and Genesis beyond Horndeski*, *JCAP* **08** (2017) 038 [arXiv:1705.06626] [inSPIRE].

[88] G. Ye and Y.-S. Piao, *Implication of GW170817 for cosmological bounces*, *Commun. Theor. Phys.* **71** (2019) 427 [arXiv:1901.02202] [inSPIRE].

[89] G. Ye and Y.-S. Piao, *Bounce in general relativity and higher-order derivative operators*, *Phys. Rev. D* **99** (2019) 084019 [arXiv:1901.08283] [inSPIRE].

[90] A. Ilyas, M. Zhu, Y. Zheng, Y.-F. Cai and E.N. Saridakis, *DHOST Bounce*, *JCAP* **09** (2020) 002 [arXiv:2002.08269] [inSPIRE].

[91] A. Ilyas, M. Zhu, Y. Zheng and Y.-F. Cai, *Emergent Universe and Genesis from the DHOST Cosmology*, *JHEP* **01** (2021) 141 [arXiv:7993129582] [INSPIRE].

[92] M. Zhu and Y. Zheng, *Improved DHOST Genesis*, *JHEP* **11** (2021) 163 [arXiv:7293198744] [INSPIRE].

[93] Y. Cai and Y.-S. Piao, *Higher order derivative coupling to gravity and its cosmological implications*, *Phys. Rev. D* **96** (2017) 124028 [arXiv:2494192924] [INSPIRE].

[94] Y. Cai, Y.-T. Wang, J.-Y. Zhao and Y.-S. Piao, *Primordial perturbations with pre-inflationary bounce*, *Phys. Rev. D* **97** (2018) 103535 [arXiv:2493194.0.] [INSPIRE].

[95] M. Zhu, A. Ilyas, Y. Zheng, Y.-F. Cai and E.N. Saridakis, *Scalar and tensor perturbations in DHOST bounce cosmology*, *JCAP* **11** (2021) 045 [arXiv:7296192553] [INSPIRE].

[96] Y. Cai, J. Xu, S. Zhao and S. Zhou, *Perturbative unitarity and NEC violation in genesis cosmology*, *JHEP* **10** (2022) 140 [arXiv:7794122447] [*Erratum ibid.* **11** (2022) 063] [INSPIRE].

[97] S. Akama and S. Hirano, *Primordial non-Gaussianity from Galilean Genesis without strong coupling problem*, *Phys. Rev. D* **107** (2023) 063504 [arXiv:7722199566] [INSPIRE].

[98] S. Mironov and A. Shtennikova, *Stable cosmological solutions in Horndeski theory*, arXiv:7727195768 [INSPIRE].

[99] P. Pavlović and M. Sossich, *Creation of wormholes during the cosmological bounce*, *Eur. Phys. J. C* **83** (2023) 235 [arXiv:77291902.7] [INSPIRE].

[100] A. Ganz, P. Martens, S. Mukohyama and R. Namba, *Bouncing Cosmology in VCDM*, arXiv:7727125802 [INSPIRE].

[101] X. Gao, M. Lilley and P. Peter, *Non-Gaussianity excess problem in classical bouncing cosmologies*, *Phys. Rev. D* **91** (2015) 023516 [arXiv:2.901.223] [INSPIRE].

[102] X. Gao, M. Lilley and P. Peter, *Production of non-gaussianities through a positive spatial curvature bouncing phase*, *JCAP* **07** (2014) 010 [arXiv:2.9514386] [INSPIRE].

[103] Y.-B. Li, J. Quintin, D.-G. Wang and Y.-F. Cai, *Matter bounce cosmology with a generalized single field: non-Gaussianity and an extended no-go theorem*, *JCAP* **03** (2017) 031 [arXiv:2027197950] [INSPIRE].

[104] S. Akama and T. Kobayashi, *Generalized multi-Galileons, covariantized new terms, and the no-go theorem for nonsingular cosmologies*, *Phys. Rev. D* **95** (2017) 064011 [arXiv:2492197370] [INSPIRE].

[105] C. Lin, J. Quintin and R.H. Brandenberger, *Massive gravity and the suppression of anisotropies and gravitational waves in a matter-dominated contracting universe*, *JCAP* **01** (2018) 011 [arXiv:2422129.47] [INSPIRE].

[106] A.P. Bacalhau, N. Pinto-Neto and S. Dias Pinto Vitenti, *Consistent Scalar and Tensor Perturbation Power Spectra in Single Fluid Matter Bounce with Dark Energy Era*, *Phys. Rev. D* **97** (2018) 083517 [arXiv:2490196659] [INSPIRE].

[107] S. Akama and T. Kobayashi, *General theory of cosmological perturbations in open and closed universes from the Horndeski action*, *Phys. Rev. D* **99** (2019) 043522 [arXiv:2629192605] [INSPIRE].

[108] S. Akama, S. Hirano and T. Kobayashi, *Primordial non-Gaussianities of scalar and tensor perturbations in general bounce cosmology: Evading the no-go theorem*, *Phys. Rev. D* **101** (2020) 043529 [arXiv:2396129005] [INSPIRE].

[109] P.C.M. Delgado, M.B. Jesus, N. Pinto-Neto, T. Mourão and G.S. Vicente, *Baryogenesis in cosmological models with symmetric and asymmetric quantum bounces*, *Phys. Rev. D* **102** (2020) 063529 [arXiv:2010.04807] [INSPIRE].

[110] P.C.M. Delgado, R. Durrer and N. Pinto-Neto, *The CMB bispectrum from bouncing cosmologies*, *JCAP* **11** (2021) 024 [arXiv:2108.06175] [INSPIRE].

[111] B. van Tent, P.C.M. Delgado and R. Durrer, *Constraining the bispectrum from bouncing cosmologies with Planck*, arXiv:2212.05977 [INSPIRE].

[112] J. Khoury, B.A. Ovrut, P.J. Steinhardt and N. Turok, *The Ekpyrotic universe: Colliding branes and the origin of the hot big bang*, *Phys. Rev. D* **64** (2001) 123522 [hep-th/0103239] [INSPIRE].

[113] W.G. Cook, I.A. Glushchenko, A. Ijjas, F. Pretorius and P.J. Steinhardt, *Supersmoothing through Slow Contraction*, *Phys. Lett. B* **808** (2020) 135690 [arXiv:2006.01172] [INSPIRE].

[114] J.K. Erickson, D.H. Wesley, P.J. Steinhardt and N. Turok, *Kasner and mixmaster behavior in universes with equation of state $w > \sim 1$*, *Phys. Rev. D* **69** (2004) 063514 [hep-th/0312009] [INSPIRE].

[115] S. Mironov, V. Rubakov and V. Volkova, *Superluminality in DHOST theory with extra scalar*, *JHEP* **04** (2021) 035 [arXiv:2011.14912] [INSPIRE].

[116] R. Kolevatov, S. Mironov, V. Rubakov, N. Sukhov and V. Volkova, *Cosmological bounce in Horndeski theory and beyond*, *EPJ Web Conf.* **191** (2018) 07013 [INSPIRE].

[117] S. Mironov, V. Rubakov and V. Volkova, *Bounce beyond Horndeski with GR asymptotics and $\gamma$-crossing*, *JCAP* **10** (2018) 050 [arXiv:1807.08361] [INSPIRE].

[118] S.S. Boruah, H.J. Kim, M. Rouben and G. Geshnizjani, *Cuscuton bounce*, *JCAP* **08** (2018) 031 [arXiv:1802.06818] [INSPIRE].

[119] S. Mironov, V. Rubakov and V. Volkova, *Cosmological scenarios with bounce and Genesis in Horndeski theory and beyond: An essay in honor of I.M. Khalatnikov on the occasion of his 100th birthday*, arXiv:1906.12139 [INSPIRE].

[120] S. Dyda, E.E. Flanagan and M. Kamionkowski, *Vacuum Instability in Chern-Simons Gravity*, *Phys. Rev. D* **86** (2012) 124031 [arXiv:1208.4871] [INSPIRE].

[121] R.H. Brandenberger, *The Matter Bounce Alternative to Inflationary Cosmology*, arXiv:1206.4196 [INSPIRE].

[122] Y.-F. Cai, S.-H. Chen, J.B. Dent, S. Dutta and E.N. Saridakis, *Matter Bounce Cosmology with the f(T) Gravity*, *Class. Quant. Grav.* **28** (2011) 215011 [arXiv:1104.4349] [INSPIRE].

[123] D. Baumann, P.J. Steinhardt, K. Takahashi and K. Ichiki, *Gravitational Wave Spectrum Induced by Primordial Scalar Perturbations*, *Phys. Rev. D* **76** (2007) 084019 [hep-th/0703290] [INSPIRE].

[124] J.-X. Feng, F. Zhang and X. Gao, *Scalar induced gravitational waves from Chern-Simons gravity during inflation era*, arXiv:2302.00950 [INSPIRE].

[125] N. Arkani-Hamed, H.-C. Cheng, M.A. Luty and S. Mukohyama, *Ghost condensation and a consistent infrared modification of gravity*, *JHEP* **05** (2004) 074 [hep-th/0312099] [INSPIRE].